\documentstyle[psfig]{l-aa}
\def\a85{ABCG~85{}}
\def\pislar{PDGLS{}}
\def\sapin{TTPR{}}
\def\J23{cluster~J2310--43{}}
\def\kms{km~s$^{-1}${}}
\def\nh{N${}_{\textsc h}${}}
\def\BCM{brightest cluster member{}}
\def\NE{North-East{}}
\def\SE{South-East{}}
\def\SW{South-West{}}
\def\NE{North-East{}}
\begin{document}
\thesaurus{11.03.1, 11.03.4,12.04.1, 13.25.2 }

\title{The rich cluster of galaxies \a85. II. X-ray analysis using the ROSAT
HRI\thanks{Based on ROSAT Archive data}}

\author {G.~B.~Lima Neto\inst{1, 2} \and 
  V.~Pislar\inst{1} 
  \and  F.~Durret\inst{1,3} \and  D.~Gerbal\inst{1,3}
  \and  E.~Slezak\inst{4}}

\offprints{G.B. Lima Neto, gastao@iap.fr }
\institute{
Institut d'Astrophysique de Paris, CNRS, Universit\'e Pierre et
Marie Curie, 98bis Bd Arago, F-75014 Paris, France
\and
	Observatoire de Lyon, Av. Charles Andr\'e, F-69561 St Genis Laval
Cedex, France
\and
	DAEC, Observatoire de Paris, Universit\'e Paris VII, CNRS (UA 173),
F-92195 Meudon Cedex, France
\and
    Observatoire de la C\^ote d'Azur, B.P. 229, F-06304 Nice Cedex 4,
France
}
\date{Received, 1997; accepted, 1997}

\maketitle

\begin{abstract}
We present a new X-ray analysis mainly based on ROSAT HRI data.  The
HRI spatial resolution combined with an improved wavelet analysis
method and with complementary radio and optical data provides
new results compared to a previous paper based on ROSAT PSPC data
(Pislar et al. 1997).
We use also redshift data in order to identify galaxies dynamically belonging
to the main body of the cluster and/or to superimposed substructures.

Various kinds of emission are superimposed on a mean thermal X-ray
emission due to the intra-cluster gas:\\
-- an X-ray flux excess in the centre at a scale of 15-25~kpc;\\
-- a south blob, partially generated by individual galaxies (such as
the second brightest galaxy); the mean velocity and velocity
dispersion of the galaxies located in this region are the same as
those of the cluster as a whole: it therefore does not seem to be a
bound subgroup;\\
-- West emission due to a foreground group with self-emission from
a Seyfert galaxy located at the north-west;\\
-- emission in the south-west due to inverse
Compton emission associated to a very steep radio source (the remnant of an
active galactic nucleus).

We have examined the possibility for the central peak to be an
``unusual'' galaxy, as assumed for the central galaxy of J2310--43
(Tananbaum et al. 1997).  We conclude on the existence of a cooling
flow region, in which the presence of at least three small features
certainly related to cooler blobs is revealed by the wavelet analysis.

We have performed a pixel-to-pixel modelling of the double X-ray
emission. The large scale emission component is comparable to those
derived from by the PSPC data and the small scale one is
interpreted as a cooling-flow.  A multiphase gas model analysis leads
to a mass deposit of 50--150~M$_\odot$/yr.

\end{abstract}

\section{Introduction}

\a85 is a richness class 1 cluster of cD type (Struble \& Rood 1987),
with an optical redshift of 0.0555.  We have recently studied it
extensively (Pislar et al.  1997, hereafter \pislar), simultaneously
using ROSAT PSPC and optical data (photometry and spectroscopy); this
led to a complex view of this cluster. With the spectroscopic capabilities,  
but modest spatial resolution of the PSPC, our previous study was aimed at
modelling
the large-scale and global properties of \a85. A full description
can be found in the quoted paper.
 
%Although the PSPC has
%spectroscopic capabilities, our previous study was aimed at overall
%properties, due to the limited spatial resolution.

Tananbaum et al.  (1997, hereafter \sapin) have pointed out that the
brightest galaxy of the cluster J2310--43 has some properties that
makes it ``unusual'': ``an active nucleus without optical emission
lines and without a substantial optical continuum''.  The \a85
Brightest Cluster Member (BCM) appears to us very similar to their
``unusual" galaxy, from the optical, X-ray and radio points of view.

The  ROSAT HRI (High Resolution Imaging) is best suited for observing small 
features in X-ray images of clusters, specially when performing a wavelet 
analysis. 

In this paper, we address various questions that we are now able 
to answer:\\
-- On the PSPC image we have found that an intense peak of X-ray emission
is superimposed on the location of the central cD, and we have shown that
this peak was \textit{unresolved}. This has led us to perform a more
detailed analysis (wavelet, spectral and modelling) on the cD of \a85
using the HRI image, and to compare its properties with those of
J2310--43.\\
-- We have shown (\pislar) that the South Blob of \a85 
appears in fact due to self-emitting X-ray galaxies.  We will check if 
this is still true with the better spatial resolution of the HRI.\\
-- Prestwich et al.~(1995) have shown the possible existence of small bright 
emitting features (of sizes 5--10 arcsec, or $8-16~h_{50}^{-1}$~kpc).
They claim that these features cannot be
identified with single galaxies and are not likely to be foreground or
background sources. We will discuss whether an ``objective" wavelet
analysis confirms the existence of the same features.\\
-- We have completed the X-ray data with optical and radio data found in the
literature, in order to check if the X-ray features detected at different
scales were correlated with other optical and/or radio objects
(extended or not). These correlations could give us a clue on the
physical nature of the X-ray features (thermal or not, for instance).

Coupling the high resolution data from the HRI with the spectral
information of the PSPC, as well as with optical and radio data, can
lead us to a better understanding of \a85.
In section \ref{data} we describe the data used for the image 
analysis (section \ref{analysis}) and spectral analysis (section 
\ref{spectre}).  The modelling of a multi-phase cooling flow and the 
detailed comparison of the \a85 and J2310--43 BCMs are done in 
section \ref{imagerie}.

\section{The data}\label{data}

\begin{figure*}[tbp]
\centerline{\psfig{figure=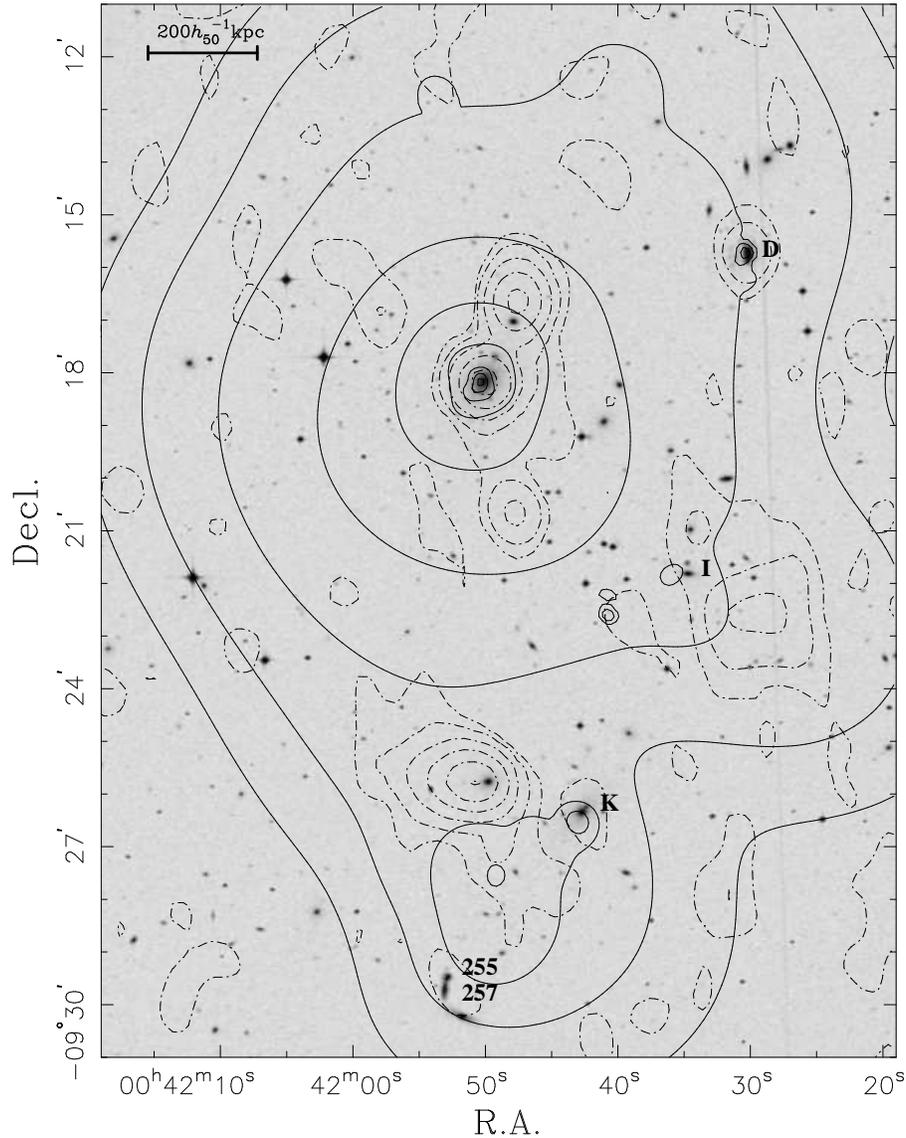,width=12cm}}
\caption{\a85 field. The grey scale image is photographic data
obtained using the UK Schmidt Telescope obtained with the
Digitized Sky Survey. The solid
contours are from the wavelet reconstructed HRI image.
The dash-dotted contours are from the VLA survey at 1400 MHz obtained from
the SkyView survey analysis. The contour levels are in logarithmic scale.
The coordinates are J2000.}
\protect\label{ondelettes}
\end{figure*}

\subsection{X-rays}

The \a85 field was observed by the ROSAT HRI in June 1992 with an exposure
time of 17351 seconds (P.I. Prestwich).  Pixels were rebinned to a size of
5 arcsec, corresponding to $8.0~h_{50}^{-1}$~kpc at the cluster 
distance, or to a size of 8 arcsec corresponding to $12.8~h_{50}^{-1}$~kpc.
In order to compare its properties with
those of the cluster J2310--43, we have also performed a new spectral analysis
based on PSPC data.

\subsection{Optical}

In the optical, our data is based on a photometric catalogue of 4232
galaxies (Slezak et al. 1997) and on a catalogue of 550 redshifts
(Durret et al. 1997). The superposition of the galaxies belonging to
\a85 (in the velocity range 13300--20000 km/s, cf. \pislar) to the
PSPC image has shown the existence of X-ray emitting galaxies, among
which the cD galaxy. The main features of the galaxy distribution were
described in \pislar.  We have found for the cluster itself
(i.e. without the fore and background substructures) a velocity
dispersion of $\sigma$=760~\kms, as expected from a richness class~1
cluster and from the X-ray luminosity of the diffuse component $\sim
(9.3 \pm 0.2)~10^{44}$~erg~sec$^{-1}$.

We have also used photographic data from the Digitized Sky Survey
obtained with the UK Schmidt Telescope to superimpose the optical images of
the galaxies to the X-ray and radio maps.

\subsection{Radio}

The correlation (or anti-correlation) of radio and X-ray emissions
can give some clues on the physical processes involved in the
cooling of the gas. It can also give limits on the magnitude
of the cluster magnetic field.

A number of radio data can be found in the literature for \a85. We present 
here a summary of these observations:
\begin{itemize}
\item A radio map at 327 MHz obtained by Kapahi and Subrahmanyan 
(see Swarup 1984) with a 3 arcmin gaussian beam;
\item VLA observations of radio sources at 1452~MHz in the direction of \a85 
(O'Dea and Owen 1985) at 1--5 arcsec resolution;
\item Data from the VLA survey at 1400~MHz with 45 arcsec resolution
performed by Condon et al. (1996);
\item Several 1500~MHz maps from the VLA Survey of rich clusters of galaxies 
with 15 arcsec FWHM (Slee et al. 1996);
\item The 2695~MHz Effelsberg map (Waldthausen et al. 1979 
and Andernach et al. 1986) made with a 4.4 arcmin beam;
\item A 4872~MHz VLA map at a resolution of 1--2 arcsec (Burns 1990). 
\end{itemize}

We will use the radio image produced by Condon et al. (1996) and compare it
to the HRI data of \a85. This radio image is available in electronic form, has a
high surface-brightness sensitivity, enough spatial resolution for our
purposes, and covers the whole field of view of the cluster.

\section{Image analysis}\label{analysis}

For the image analysis, we have performed a wavelet multiscale 
reconstruction on the HRI image.  This multiscale analysis is 
explained and described in \pislar. However, we have improved our 
method by correcting the image for vignetting (using the Snowden 1995 
``cookbook''), but not for background contribution, since we 
need to know the background in order  to determine the 
significance level of the features used for the final wavelet 
``reconstructed'' image. We have modified the wavelet technique, in order to
detect better the faint background components which are sometimes hidden 
by bright small-scale superimposed features.  Details on the recent 
improvements to this method are given in the Appendix.

The HRI has a better spatial resolution than the PSPC -- the pixel size
is 5 arcsec instead of 15 arcsec -- therefore,
although the observed cluster is obviously the same for the two detectors,
a comparison of the two analyses can be fruitful. Actually, we have paid
attention to three kinds of behaviours that we will describe below.

\subsection{Comparison of PSPC and HRI images}\label{PSPC/HRI}

We will compare the wavelet ``reconstructed'' HRI
(Fig.~\ref{ondelettes}) and PSPC images (Fig.~2 of \pislar). In
addition to the X-ray image, we overlay in Fig.~\ref{ondelettes} a
1400~MHz radio map and an optical image.

When doing the overlay with the optical image, we noticed that the
X-ray HRI contours were shifted towards the east by about 9 arcsec,
when comparing the X-ray sources related with galaxies with
the positions of the galaxies D, K, and the cD (cf. Table 1 of
\pislar). The same shift is also seen between the X-ray and radio
emissions of these same galaxies. This shift is actually comparable to
the upper limit of the known uncertainties of the ROSAT aspect camera.
Therefore, in Fig.~\ref{ondelettes} (as well as Figs.~\ref{southblob}
and \ref{ondes} below) we applied a 9 arcsec correction to the
position of the X-ray image, in order to realign it with the optical
and radio images.

The PSPC and HRI general image structures are essentially the same:
a main body with an intense peak superimposed (this prominent X-ray peak
being centered on the location of the cD), two extensions to the
west and a south blob. However, some differences can be noticed as
described in the paragraphs below.

\subsubsection{Central emission}\label{central}
The central peak is now \textit{resolved} at the 10 arcsec scale
(plane 2 of the wavelet coefficients), but remains \textit{unresolved}
at the 8 arcsec scale. It is definitively \textit{not} a point-like
source. This suggests that the scale of the summit is greater
than 13 $h_{50}^{-1}$ kpc but smaller than 16 $h_{50}^{-1}$ kpc.

\begin{figure}[tbp]
	\centerline{\psfig{figure=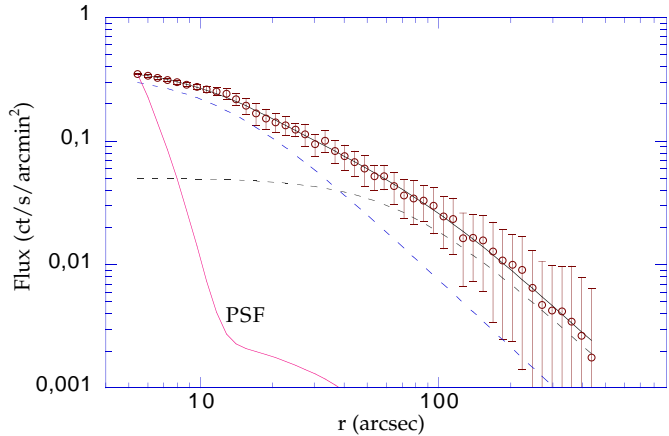,width=8.8cm}}
	\caption{HRI X-ray brightness profile of \a85.
The dashed lines are the central and extended components
obtained by fitting two $\beta$-models to the brightness. The continuous line
is the sum of the two components. The dotted line is the convolution
of a point-like emission with the HRI Point Spread Function}
\protect\label{profile}
\end{figure}

We give in Fig.~\ref{profile} the profile obtained using the IRAF
stsdas/ellipse task. A small shoulder is visible at a radius of about 15
arcsec (24 $h_{50}^{-1}$ kpc); this is also the sign of an X-ray
emitting central source superimposed on an extended emission.

We show also in Fig.~\ref{profile} that the brightness profile may
be modelled by the sum of two components, each one following
a $\beta$-model. This is only a rough fit since we do not take
into account the physical and geometrical properties of the gas.
For comparison purposes we show a point-like source convolved
with the Point Spread Function of the HRI.
A more accurate fit and physical discussion of the two components
will be made in \S~\ref{untangle}.

%%%%%%%%%%%%%%%%%%%%%%%%%%% Figure %%%%%%%%%%%%%%%%%%%%%%%%%
%%%%%%%%%%%%%%%%%%%%%%%%%%%%%%%%%%%%%%%%%%%%%%%%%%%%%%

\subsubsection{The south blob}
The south blob includes at least three bright spots. However, there is
 an important difference between the PSPC and HRI images:~in \pislar\
 the diffuse X-ray emission visible on the ``reconstructed'' image of
 the south blob disappears completely when the reconstruction is
 processed without taking into account the smallest scale structures.
 This is why we have concluded that:
\begin{quote} ``the south extension S is constituted by the 
superposition of small emission regions and not by a diffuse extended 
source such as hot X-ray emitting gas in a group of galaxies''.  
\end{quote}
On the contrary, in the HRI data, when small scale features are 
removed a faint diffuse component remains.  This is perhaps due to 
the fact that the lower spatial resolution of the PSPC blurs images 
and makes the faint diffuse emission of the south blob undetectable.  
One should also take into account the better performance of the 
improved wavelet image reconstruction that we have used (see the   
Appendix).
 
\begin{figure}[tbp]
	\centerline{\psfig{figure=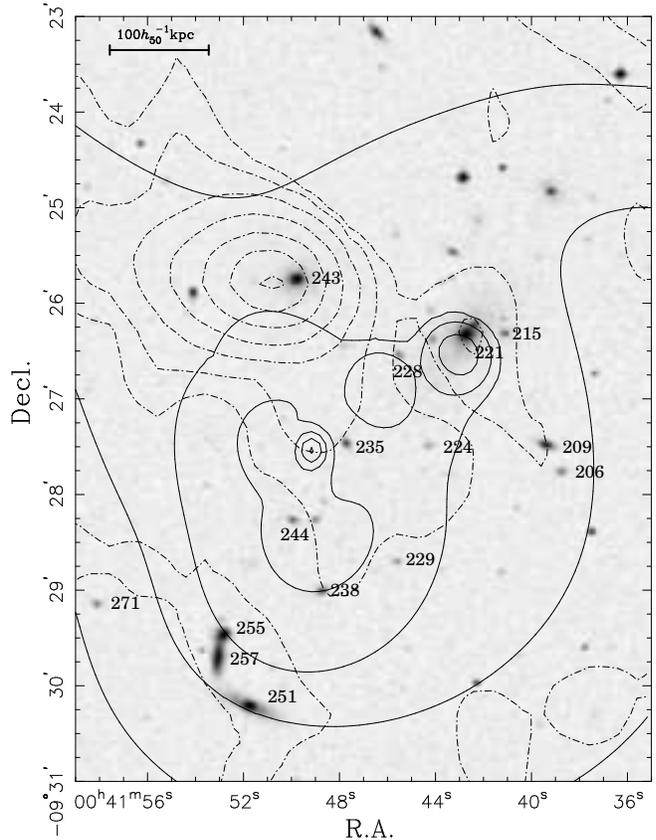,width=8.6cm}}
	\caption{South blob (reconstructed HRI image and VLA radio) with the 15
galaxies from Table~\ref{gal} superimposed. The galaxies which are
not marked are either in the foreground or background. The contour lines have the
same meaning as if Fig.~\ref{ondelettes}}
\protect\label{southblob}
\end{figure}

  We have superimposed 15 known galaxies with velocities between 14000
  and 19000~km/s in the region of the south blob as shown in
  Fig.~\ref{southblob}.  Details on these galaxies are given in
  Table~\ref{gal}, a subset of the complete catalogue of the \a85
  field (Durret et al.  1997).

%%%%%%%%%%%%%%%%%%%%%%%%%%% Figure %%%%%%%%%%%%%%%%%%%%%%%%%
%%%%%%%%%%%%%%%%%%%%%%%%%%%%%%%%%%%%%%%%%%%%%%%%%%%%%%%%%%%%%%%%%%%%%%

%%%%%%%%%%%%%%%%%%%%%%%%%%% table %%%%%%%%%%%%%%%%%%%%%%%%%
\begin{table}[htbp]
	\centering
	\caption{Galaxies in the south blob (velocities
between 14000 and 19000~km/s, R~$<18.5$) }
	\begin{tabular}{lcccc}
		\hline
Galaxy &$ \alpha$ & $\delta$ & Velocity& R\\
       & (J2000.0)& (J2000.0)& (km/s) & (magnitude)\\      
		\hline \\
206& 0 41 38.94 & -9 27 47.91 &17216  &17.6\\
209& 0 41 39.54 & -9 27 31.27 &16647  &16.2\\
215& 0 41 41.34 & -9 26 21.31 &16357  &17.3\\
221 (K) & 0 41 43.05 & -9 26 22.32 &16886  &14.1\\
224& 0 41 44.54 & -9 27 31.41 &15517  &18.3\\
228& 0 41 45.82 & -9 26 34.70 &15033  &17.8\\
229& 0 41 45.92 & -9 28 44.50 &16168  &18.3\\
235& 0 41 48.04 & -9 27 30.38 &17203  &17.0\\
238& 0 41 49.03 & -9 29 03.39 &18437  &17.0\\
243& 0 41 50.24 & -9 25 47.41 &17360  &15.0\\
244& 0 41 50.34 & -9 28 18.32 &17081  &17.4\\
251& 0 41 52.22 & -9 30 16.43 &17164  &14.6\\
255& 0 41 53.23 & -9 29 29.45 &15751  &15.8\\
257& 0 41 53.62 & -9 29 45.45 &15392  &15.1\\
271& 0 41 58.64 & -9 29 10.83 &16568  &17.8\\
	 	\hline
	\end{tabular}
	\label{gal}
\end{table}
%%%%%%%%%%%%%%%%%%%%%%%%%%%%%%%%%%%%%%%%%%%%%%%%%%%%%%%%%%%%%%%%%%%%%%

%%%%%%%%%%%%%%%%%%%%%%%%%%% figure %%%%%%%%%%%%%%%%%%%%%%%%%%%
\begin{figure}[tbp]
	\centerline{\psfig{figure=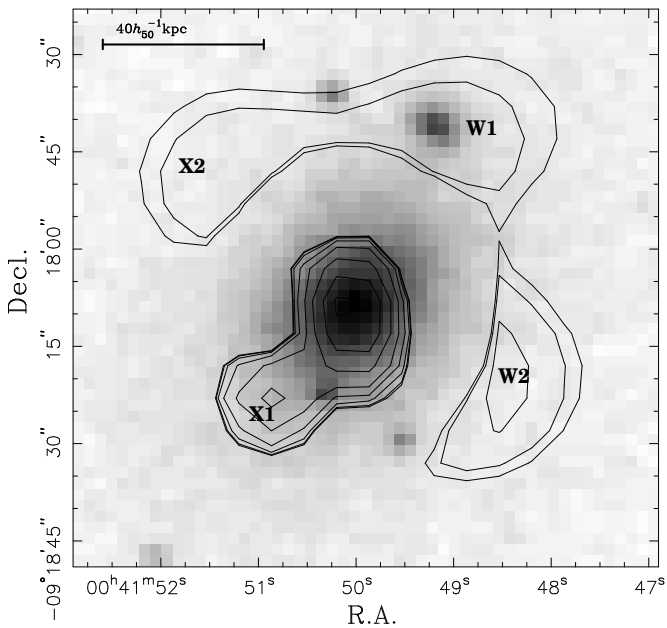,width=8.8cm}}
	\caption{Wavelet coefficients at a scale of 5 arcsec (one HRI pixel)
	superimposed on the optical image}
\protect\label{ondes}
\end{figure}
%%%%%%%%%%%%%%%%%%%%%%%%%%%%%%%%%%%%%%%%%%%%%%%%%%%%%%%%%%%%%%%%%%%%%%

The most luminous galaxy in this region  (galaxy 221 in Table~\ref{gal},  
hereafter noted K as in \pislar, with a magnitude R~=~14.1)
 is actually the second most luminous galaxy belonging to
the central main body of \a85. 
%The use of the ``hierarchy clustering method'' 
%(Serna \& Gerbal 1997), indicates that in fact it forms a bound pair with the
%central cD with a projected distance of about 800 $h_{50}^{-1}$ kpc.
This luminous galaxy is located in a local potential well, generated 
by the galaxy itself as well as by some other bright galaxies in the vicinity 
(including galaxy 251 with R~=~14.6), in the same way that the cD lies 
at the bottom of the central gravitational well.  Remembering that the 
X-ray surface brightness traces the potential, such a gravitational 
well would be sufficient to generate the diffuse X-ray component of 
the south blob.

However, contrary to Kneer et al. (1996), we do not find evidence that the
south blob is falling into the main cluster.  The mean and the median
velocities of the 15 galaxies superimposed on the south blob
(Table~\ref{gal}) are 16585 and 16650 km/s respectively, which are
close to the mean and median velocities of the whole cluster. The
velocity dispersion of these 15 galaxies is $900\pm 200$~km/s, which
is comparable, within the error bars, to that of the whole cluster, 
but is too high for a subgroup velocity dispersion.
Notice that the presence of these 15 galaxies only means a galaxy
over-density inside the cluster, but not the presence of a
self-gravitational bound substructure.

We believe in the dynamical explanation 
proposed above, that the south blob is an X-ray over-brightness response to
the gravitational well generated by the second
\BCM. The fact that the orientation of this galaxy is roughly the
same as the main orientation of the blob itself gives support to this point
of view.

\subsection{Central features}

Prestwich et al. (1995) have looked in the central part of \a85; they 
have suggested that some
spots are not related to X-ray emitting galaxies.  This has been done 
by subtracting a large scale emission model from the true image.  But 
finally only one of these spots (the feature noted X1 in Fig.~3 of their 
paper) has been considered as significant by the authors, while the 
second spot X2 that they found has been rejected as unsignificant.

A picture of the wavelet coefficients at scales of 1 pixel is
displayed in Fig.~\ref{ondes} for the same central region. As
mentioned above, a shift of about 9 arcsec of the HRI center has been
done towards the west (see \ref{PSPC/HRI}), so that the main central
feature coincides exactly with the cD galaxy.

We remind the reader that with the wavelet technique only features
significant at a 3$\sigma$ level are taken into account.  X1, X2, W1
and W2 are significant features found on our map.  W2 (as well as 
X1 and X2) has no
optical counterpart, while W1 does have one (when we take into
account the error on the HRI attitude). However, we have no
velocity information on the galaxy associated with W1 and we
cannot be sure that it is not a fore or background galaxy.

We thus confirm the existence of at least three bright spots related to
inhomogeneities of the cooling flow.  The bright spots appear
distributed with a spherical symmetry around the central galaxy.
Obviously, the gas in the central region, which is probably in the
cooling flow regime, is not homogeneously distributed (both the
density and temperature must be inhomogeneous, according to Prestwich
et al. 1995).

\subsection{Radio and X-ray comparison}

%%%%%%%%%%%%%%%%%%%%%%%%%%%%%%%%%%%table %%%%%%%%%%%%%%%%%%%%%%%%%%%%%%%%%%%
\begin{table*}[htbp]
	\centering
	\caption{Spectral analysis results}
	\begin{tabular}{ c c c c c c c}
	\hline
$T$        & \nh        &     Z     & $\alpha $& L$_{\textrm X}$&$\chi^2/DOF$& Case \\
(keV)       &($10^{20}$cm$^{-2}$)&(Z$_\odot$)  &    &(10$^{44}$~erg~s$^{-1}$)& &\\
\hline
$4.6\pm 1.2$ &$2.6\pm 0.3$&$0.25\pm0.3 $&   ---       &9.0   &237/207& 1\\
\\
3.1$\pm 0.2$    &  canonical & $< 0.1$&    ---      &9.0   &334/208  &2\\
\\
4.3$\pm 2.0$ &            &0.24$\pm0.3 $&   ---       &8.6 &        &3\\
0.10$\pm0.03$&  canonical &0.3 fixed    &   ---       &1.3 &236/206 &3\\
\\
5.1$\pm 1.1$ &  canonical &0.3 fixed    &          &8.0   & &4\\
             &            &             &-3.25$\pm0.7$&2.6& 234/207   &4\\
\hline
	\end{tabular}
	\label{spectral}

Note: The error bars are correlated $3\sigma$ uncertainties.
\begin{enumerate}
\item  Temperature, hydrogen column density, \nh, and metallicity, Z, are free
parameters in a Raymond-Smith (R--S) plasma. These values have been discussed
in \pislar.
\item  Because the previous value for \nh\ is lower than the
``canonical" value -- the galactic value of 3.58~10${}^{20}$ cm${}^{-2}$,
as given by Dickey \& Lockman (1990) in the position of the cD --
we have fixed \nh\ to be
equal to this canonical value in the fitting process. However the $\chi^{2}$
is bad.
\item  If we fit the spectrum with two R--S laws, and fix \nh\ and Z, notice
the very weak temperature of the cooler component of $\sim 0.1$keV.
The $\chi^{2}$ is then improved.
We have also tried other fixed values for the metallicity, but due to the bad
quality of the spectral capability of the PSPC the resulting temperatures do
not change significantly.
\item If we fit the spectrum with a R--S and a power law, the R--S
temperature is the same as in the other cases (within error bars), and the
power law index is very steep.
\end{enumerate}
\end{table*}
%%%%%%%%%%%%%%%%%%%%%%%%%%%%%%%%%%%%%%%%%%%%%%%%%%%%%%%%%%%%%%%%%%%%%%

In Fig.~\ref{ondelettes}, we have presented the isocontours of the wavelet
reconstructed X-ray HRI image, superimposed on the optical image and
radio contours at 1400~MHz.

%%%%%%%%%%%%%%%%%%%%%%%%%%%
\begin{figure}[tbp]
	\centerline{\psfig{figure=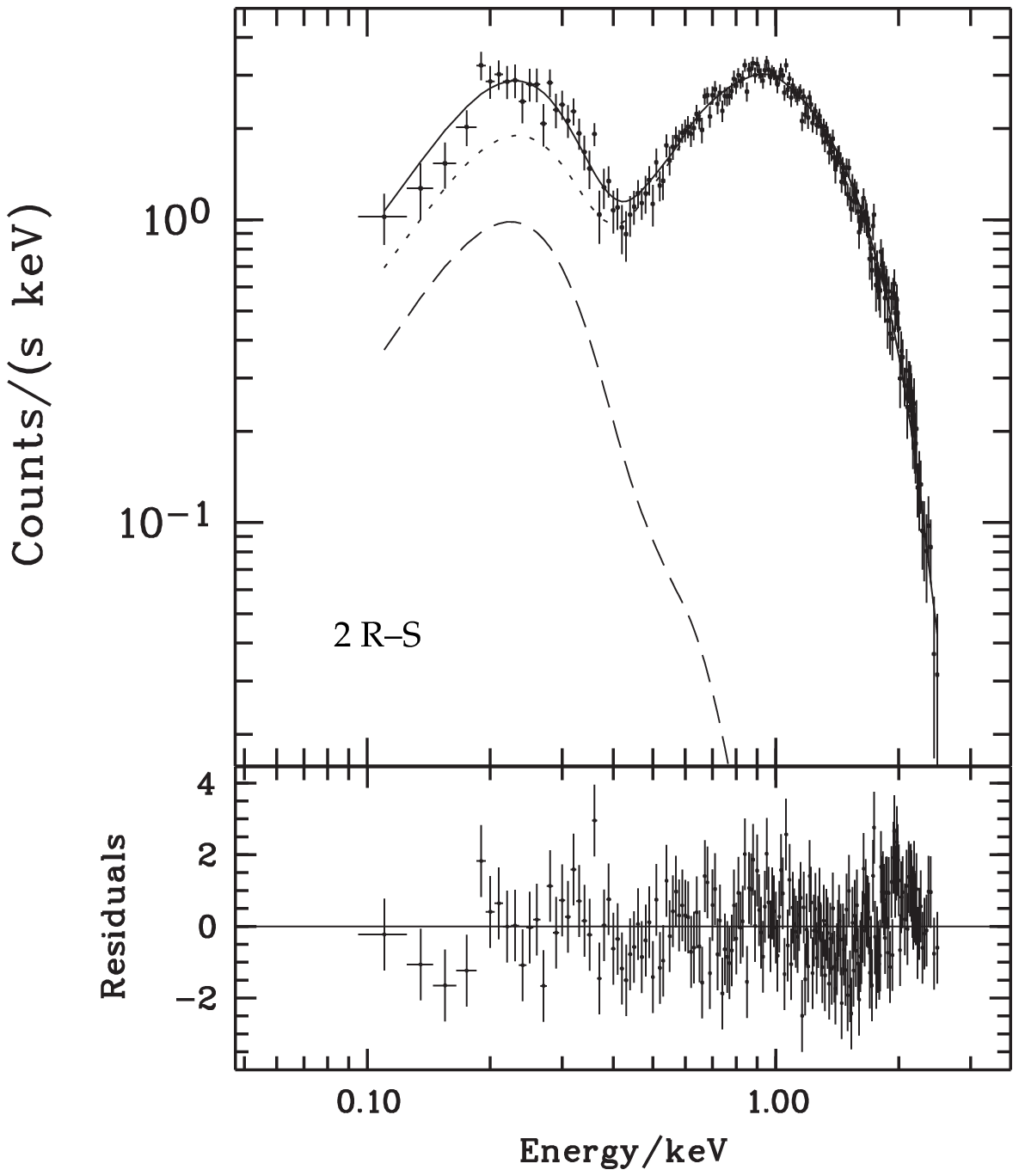,width=7.8cm}}
	\centerline{\psfig{figure=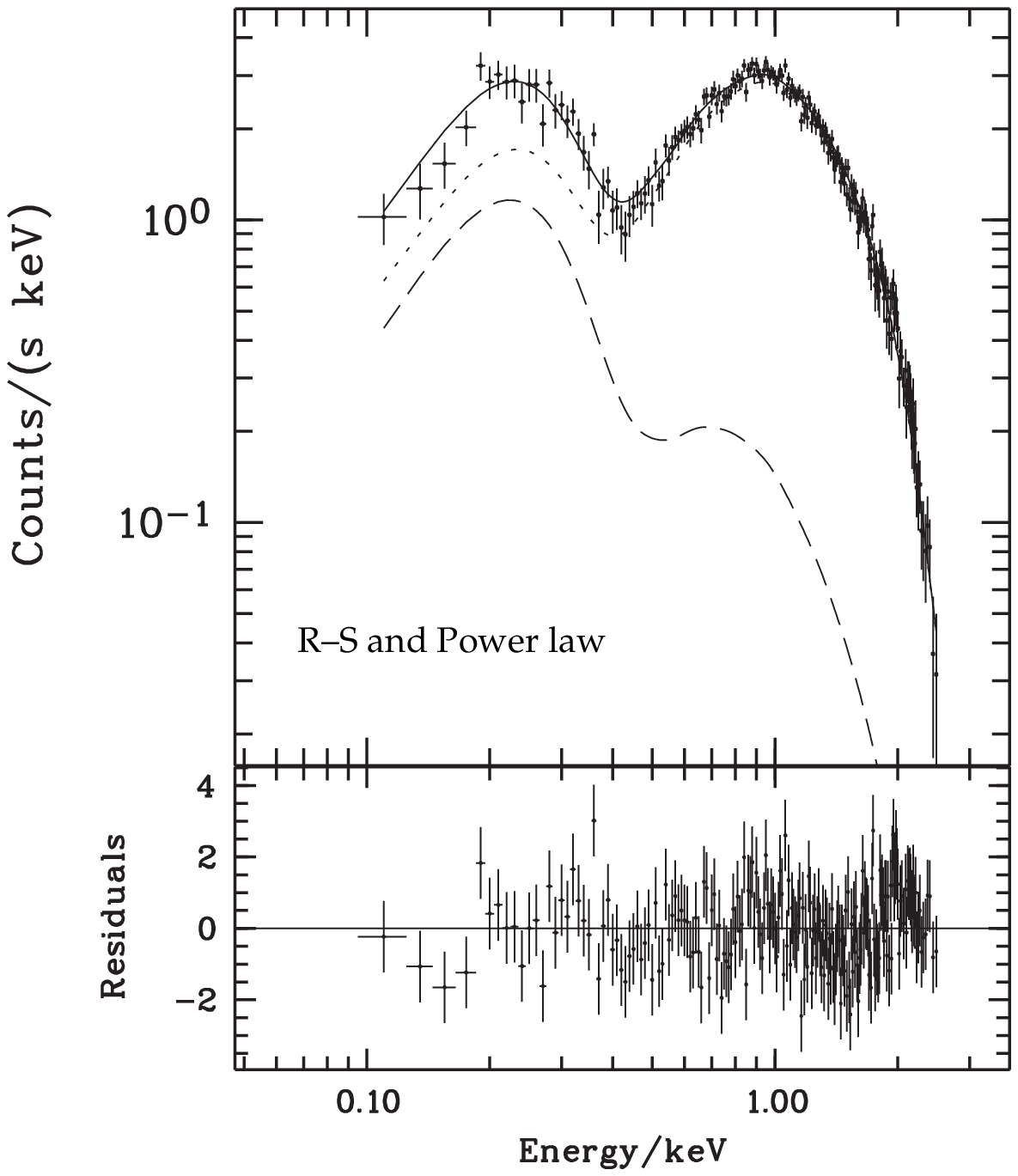,width=7.8cm}}
	\caption{Background subtracted pulse height spectra. Top: two R--S.
 Bottom: R--S plus a power law}
\protect\label{fspectre}
\end{figure}
%%%%%%%%%%%%%%%%%%%%%%%%%%%%%%%%%%%%%%%%%%%%%%%%%%%%%%

At the position of the cD galaxy, we see strong radio emission. It
is also the case around galaxies D and K.  \NE\ of K we see
radio emission around galaxy 243 and an extension of X-ray emission
from the south blob. These radio emissions, associated with both X-ray
and optical counterparts, may suggest the presence of some galactic
nuclear activity. Thus, part of the X-ray emission may be due to a non
thermal source.

\SE\ of K, we see a couple of galaxies likely to be interacting
(257 and 255) in the centre of a small radio emission, also correlated
with an excess of X-ray emission.

We also see strong radio emission with no optical counterpart at 
the \SW\ of galaxy I, corresponding to an extension of the X-ray 
emission (which is also visible with the PSPC).  This radio emission 
is actually due to the very steep spectrum radio source MRC~0038-096.  
Swarup (1984) has detected this source at 327 MHz, and it is also visible 
by Slee et al. (1996) at 1500 MHz; it is barely detected at 2700 MHz 
by Andernach et al. (1986), but not visible by Waldthausen et al.  
(1979) at the same frequency.  MRC~0038-096 is possibly a relic source 
(Goldshmidt \& Rephaeli 1994, Komissarov \& Gubanov 1994), the remnant 
lobes of an once active radio galaxy. In this case, the radio 
emission is accompanied by an X-ray emission produced by inverse 
Compton scattering of the cosmological microwave background.

\section{Spectral properties}\label{spectre}

In \pislar\ we have paid attention essentially to the general spectral 
properties, and have also searched for possible variations of the 
temperature, metallicity, and hydrogen column density,
\nh, with radius.  The actual existence 
of the central emission peak hinted by our wavelet analysis and the 
X-ray brightness profile lead us now to look for a double structure in 
the spectral properties.  This will also allow a comparison with the 
properties of the brightest member of the cluster J2310--43 
(\sapin). 

 Notice that in their case the central galaxy is hosted in a very poor
 cluster (richness class zero), with the bulk of the emission coming
 from the central galaxy only, so \sapin\ attempted to fit the
 spectrum with a single power law.  This is not the case for \a85,
 where an important emission from the overall cluster is observed.
 Therefore, even if we suppose that the central emission in the
 central part of \a85 really corresponds to an ``unusual" powerful
 galaxy, the main difficulty would be to untangle the spectral and
 geometrical properties of the cD: there must at least be two
 components.

%%%%%%%%%%%%%%%%%%%%%% table %%%%%%%%%%%%%%%%%%
\begin{table*}[htbp]
	\centering
	\caption{Modelling results of the pixel--by--pixel fitting of the HRI data}
	\begin{tabular}{l c c c c c c }
		\hline
$N_{0}$           &$\beta$           &$r_{c}$   &Temp. & $X^2$& Notes \\
($10^{-3}$cm$^{-3}$)&                 &(kpc)    &(kev) &    &   \\
\hline
$5.0\pm 0.7$    &$0.464\pm 0.013$  &$41\pm 6$&4 &           12929  & 1 \\
\\
$4.2\pm 0.5$  &$0.475\pm 0.015$  &$49\pm 8$&4 &          12812  & 2 \\
%   $9.9\pm 2.6$  &$0.36\pm 0.01$   &$13\pm 4.5$&4&          8398 & 3 \\
%                  &                  &            &$146\pm 23$&     &3  \\
\\
$1.1\pm 0.2$  &$0.60\pm 0.05$  & $191\pm 38$ &4.0 &         12456  & 3\\
$45.0\pm 7.0$  &$0.51\pm 0.05$  & $27\pm 6$ &0.1&         & 3\\
		 		\hline
	\end{tabular}
	\label{ajustements_hri}

Note: The error bars are uncorrelated $3\sigma$ uncertainties,
obtained directly from the principal diagonal of the error matrix.
\end{table*}
%%%%%%%%%%%%%%%%%%%%%%%%%%%%%%%%%%%%%%%%%%%%%%%%%%%%%%

Therefore using the EXSAS package in MIDAS we have redone various
kinds of spectral analyses with the PSPC data.  The results are
summarized in Table~\ref{spectral}.

 The two component fitting corresponds to cases 3 \& 4 in
 Table~\ref{spectral}; we give in Fig.~\ref{fspectre} the pulse height
 spectra corresponding to these cases.  Notice the very similar
 aspects of the two spectra, the first one corresponding to the low
 temperature Raymond--Smith model (top--case 3) and the second one to
 a power law (bottom--case 4).  Both values of $\chi^{2}$ are
 practically the same, a clear confirmation that it is not possible to
 discriminate the actual physical process.

\section{Modelling}\label{imagerie}

The wavelet analysis of the HRI image leads to the following view:~a 
strong emission corresponding to the cD galaxy position is superimposed on 
large scale emission obviously corresponding to the cluster itself.

\subsection{Untangling the cD emission from that of the cluster}\label{untangle}

To overcome the problem of superposition of the cD on the cluster,
we have tried various kinds of modelling. Notice that the ``optical" cD galaxy
is very large and thus its emission is not necessarily point-like with the
HRI resolution.

We use a pixel--by--pixel maximum likelihood fitting procedure, taking
into account all the properties of the PSPC or HRI (see a full
description in \pislar). In order to increase the signal to noise
ratio, the data were binned to 10'' pixels.  The number of pixels is
17101 corresponding to a limiting radius $R_{L}=$~13 arcmin; this
limit is imposed by the size of the HRI field.  We have used the data
after processing them following the Snowden (1995) cookbook. The
$\chi^2$ statistics do not apply because of the low number of counts
per pixel, so we used the more general maximum likelihood technique, assuming
that the count distribution is poissonian.

The results are summarized in Table~\ref{ajustements_hri}.  In view of 
comparing models to one another, we have calculated the ``distance" 
(noted $X^{2}$) between the observed 
number of photons and the synthetic number of photons per pixel.

These results raise the following comments (the numbers correspond
to the ``Notes'' column in Table~\ref{ajustements_hri}):

\begin{enumerate}
\item  In view of comparing the PSPC (see \pislar) and HRI images of \a85, we have
fit a unique $\beta$-model; we have found small values for $\beta$ and
for the core radius, comparable to the values based on PSPC data.
\item We have added to a $\beta$-model
a point-like structure i.e. the point-spread-function as
given in the HRI Calibration Report (David et al. 1996);
$X^2$ is then improved.
The luminosity of the point-like component relatively to the main component
is $\sim 5.4\ 10^{-3}$ (up to 13 arcsec).
\item We have used the sum of two $\beta$-models. One is supposed to
account for the large scale structure and the other for the cD.  The
larger component is probably due to the diffuse
component (the overall cluster).  The second $\beta$-model would then
really correspond to the cooling flow region, with a very small core radius 
and a higher density. The luminosity ratio between the inner and outer components
is $\sim 0.5$ (up to 13 arcsec).

\end{enumerate}

The best fit is obtained with two $\beta$-models. When computing the
emissivity profiles corresponding to this fit, we note that the
larger and hotter component becames dominant at a radius of about
150~$h_{50}^{-1}$~kpc.

\subsection{Comparison of the central galaxies of \a85 and \J23}\label{tanenbaum}
We first indicate the main general properties of the \a85 cD compared
to those of the central galaxy in \J23.  The main properties of these
two galaxies, as derived from \sapin, from the literature and from our
analysis are indicated in Table~\ref{comparaison}.  The item number
corresponds to that in the ``remarks'' column.

%%%%%%%%%%%%%%%%%%%%%%%%%%%%%%%% table %%%%%%%%%%%%%%%%%%%%%%%%%%%%%%%%%%%%%
\begin{table}[htbp]
	\centering
		\caption{Properties of the two \BCM{}s of \J23 and \a85}
		\begin{tabular}{ l c c c }
		\hline
		 & J2310--43 & \a85 cD& remarks  \\
		\hline
		Richness Class & 0 & 1 & \\
		Redshift & 0.0886 & 0.0555&  \\
		Type & D & cD & 1  \\
		B--V & 1.19 & 1.04& 2  \\
		Power law index & $-1.4$ & $-3$ & \\
		Emission lines & no & no & 3 \\
	 	Other emission & yes & yes &4\\
		\hline
	\end{tabular}
\label{comparaison}
\end{table}
%%%%%%%%%%%%%%%%%%%%%%%%%%%%%%%%%%%%%%%%%%%%%%%%%%%%%%%%%%%%%%%%%%%%%%%%%%

\begin{enumerate}
\item The cD aspect of the \a85 BCM is determined principally because
it is a giant galaxy with an envelope, but
having a variation in the optical brightness profile slope
between the core and the outer part (Colless, private
communication).
J2310--43 is described by \sapin\ as a D galaxy but with a halo; we therefore
wonder if these two galaxies are really of different types.
\item  The (B--V) data come from \sapin\ for the central galaxy of J2310--43
and from the NED data bank for the \a85 cD.
\item  A spectrum of the central galaxy of \a85 kindly provided
by Colless prior to publication does not show characteristics of an active
galactic nucleus.
\item The observations of \a85 at radio wavelengths (e.g. Swarup 1984) show
strong emission associated with the cD (cf. also Fig.~\ref{ondelettes}).
The Parkes-MIT-NRAO survey (Griffith \&
Wright 1993) associates a radio source with J2310--43.
\end{enumerate}

The BCM galaxy of \J23 studied by \sapin\ essentially has the
properties which are normal for large ellipticals, but with a huge
X-ray emission. These authors have fit its X-ray spectrum using a power-law
with index $\sim -1.4$.  They concluded that this object is an
example of a new class of objects and by continuity related its properties
to those of BL~Lacs.

 \sapin\ have proposed various candidates which may belong
to the same class of objects they have defined. Their study is essentially
based on a PSPC observation, and they have pointed out the necessity to
perform further investigations with the HRI, which we are doing  here 
for \a85.

The power-law spectra found in section~\ref{spectre}
is very steep (slope $\simeq 3$) but is certainly an artifact in the sense
that a Raymond-Smith plasma model is also a good fit; in fact, in the PSPC
energy range it is not possible to discriminate between the two kinds
of spectra.  From the strong radio emission associated with the \a85
cD, one may suppose that there is some kind of nuclear activity in this
galaxy, so that some X-ray emission may be related to the radio
source.

However, the bulk of the central emission has a thermal origin: we
have seen that the emission is not point-like (section
\ref{PSPC/HRI}), thus ruling out quasi-stellar emission, which would be
expected for an ``unusual'' galaxy; the central X-ray
emission is extended, and is surrounded by some small X-ray emitting
sources (Fig.~\ref{ondes}); it seems reasonable that the X-ray excess
comes from a cooling flow.

\subsection{Cooling flow}

The emissivity of \a85 is well modeled by two superposed $\beta$-models,
one of large extent (the cluster) and a more powerful one of smaller
radial extent. The central emission has a dimension of $\sim 15 
h_{50}^{-1}$~kpc (section \ref{PSPC/HRI}).

We use here the multi-phase cooling flow theory introduced by
Nulsen (1986) and developed by Thomas (1988), Gunn \& Thomas (1996) and
Teyssier (1996). In this approach, we adopt a simple form of the deprojected 
emissivity profile (corrected for the instrumental response) given by:
%%%%%%%%%%%%%%%%%%%%%%%%%%%%%%%%%%%%%%%%%%%%%%%%%%%%%%%
\begin{equation}
\varepsilon(r) = \frac{43.05}{r^{3.32}}
\left(\frac{r}{r+13.8}\right)^{2.46} 
\end{equation}
%%%%%%%%%%%%%%%%%%%%%%%%%%%%%%%%%%%%%%%%%%%%%%%%%%%%%%%
with $r$ given in HRI pixels (5 arcsec). 
We use the temperature obtained with the PSPC (see \pislar)
to normalize our results and set the maximum temperature equal
to 4.5~keV (which occurs at $\sim$~220 kpc).

The multi-phase is characterized by the density distribution function in the
phase space, $f(\rho)$, that determines the contribution
of the densities to the mean density, $\overline{\rho} = \int f(\rho)
\rho d\rho$. Following Thomas (1996), we take $f(\rho) \propto
\rho^{-7/2} (1-\rho^{-3/2})^{k-1}$, where $k$ is a free parameter. We
assume a power-law cooling function $\Lambda\propto T^{0.5}$, and a
fully ionized plasma.

%%%%%%%%%%%%%%%%%%%%%%%%%%%
\begin{figure}[tbp]
	\centerline{\psfig{figure=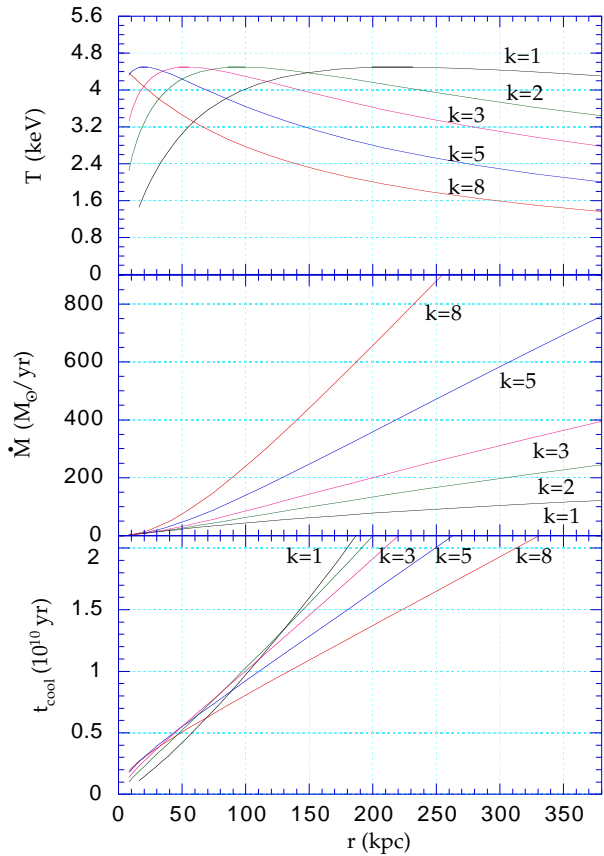,width=8cm}}
	\caption{Multi-phase cooling flow of \a85. Top, the mean temperature
	of the gas in keV;middle, the mass deposit rate in solar units per year;
bottom, the cooling time in units of $10^{10}$~years. The parameter
$k$ determines the distribution of phase of the gas (see text)}
\protect\label{coolfig}
\end{figure}
%%%%%%%%%%%%%%%%%%%%%%%%%%%%%%%%%%%%%%%%%%%%%%%%%%%%%%%

In Fig.~\ref{coolfig} we present the results of the multi-phase
cooling flow in \a85, where we focus on the region where the cooling
time $t_{cool}$, is smaller than the age of the cluster. In order to
determine the maximum radius where the cooling is effectively
important, we assume $t_{cool} \approx H_0^{-1} \approx 1.5\ 10^{10}$
yr; this corresponds to a cooling radius, $r_{cool}$, varying from 140
to 220~kpc, depending on the value of the parameter $k$.  The top
panel shows the predicted temperature profile. The temperature is not
isothermal in the multi-phase model and, after reaching a maximum,
decreases with radius (for the curve corresponding to $k=1$, the
decrease in temperature is beyond the limit shown in the figure).
Taking into account the errors in the temperature measure, the curve
may be shifted up or down by about 1~keV. For $k \ga 5$ the maximum
temperature is found very near the centre. The drop in temperature in
the outer region for these values of $k$ seems incompatible with the
PSPC observations (figures 8 and 10 from \pislar), which favours $k
\la 3$.

The mass deposit, $\dot{M}$ grows almost linearly
with radius and reaches values ranging from $\sim 50$ to
150~M${}_\odot$/yr for $1 \le k \le 3$.

Note that the above results depend on the deprojection of the 2D
emissivity profile, especially near the centre. As noticed by
Thomas (1996), a flat emissivity profile would imply higher values
of $k$. The multi-phase
model is also particularly sensitive to the temperature profile,
which could, in principle, be used to constrain the free parameter
of the multi-phase model.

\section{Discussion and conclusions}

We have continued in this paper the study of \a85 begun in Paper~I
(\pislar\ ). Two other papers (Slezak et al. 1997 and Durret
et al. 1997) are devoted to the photometric and redshift
catalogues, while a third one (Durret et al. in preparation) will be
devoted to the optical analysis.

In this paper we have made use of X-ray ROSAT HRI, optical and 
published radio data. 

Observed by the ROSAT PSPC or HRI, the mean features of \a85 are of course the 
same; but the better HRI resolution of a few arcsec (the size of a 
galaxy at the distance of \a85) combined with the PSPC X-ray 
spectroscopy, optical information (2D+1D, imaging plus redshifts), and 
radio data provides the means to go one step further in understanding 
the physics of this rich cluster of galaxies.

The bulk of the X-ray emission comes from thermal bremsstrahlung due to the 
intra-cluster gas that extends over at least 800 $h_{50}^{-1}$kpc. 
Several other sources are superimposed on this overall emission.

Using a wavelet analysis, we were able to resolve an excess of X-ray
flux in the centre at a scale of about 15--25 $h_{50}^{-1}$~kpc).  The
central activity may be due to a cooling flow; the X-ray
emissivity is indeed compatible with a multi-phase cooling flow model with a
cooling radius of about 150 $h_{50}^{-1}$~kpc and a mass deposit rate
of 50--150 M${}_\odot$/yr. This cooling flow may also be triggering
the strong central radio source activity and thus, a contribution of
X-ray emission may be due to a non-thermal process.  However this is
in contradiction with the situation of the \J23 D galaxy as proposed by
\sapin. The interpretation we give for the X-ray emission for the cD
of \a85 casts some doubt on the interpretation of the \J23 D galaxy as
``unusual'', since these two galaxies seem to be similar.
Nevertheless, high resolution observations of \J23 are needed
to verify whether or not its X-ray emission is point-like. If it is
indeed unresolved at a scale of a few kpc, then
the X-ray excess in \J23 could hardly be explained as an extended cooling-flow.

A detailed analysis of the very central part confirms the existence of
at least 3 small X-ray features (one in common with Prestwich et
al. 1995). These features are most probably the sign of an
inhomogeneous cooling flow, corresponding to cooler regions compressed
by the hotter surrounding gas.

To the south of the main component, there is a blob coinciding with a
small group of galaxies, among which the second brightest cluster
member. This group of galaxies, however, does not appear to be a bound
substructure: its velocity dispersion is too high, since it is
actually comparable to the velocity dispersion of the whole
cluster. We therefore interpret the south blob as a fortuitous and
transient overdensity of galaxies. The short relaxation time scale of
the intra-cluster gas explains the fact that we observe an X-ray
envelope around the galaxies of the south blob. Note that part of the
X-ray emission is connected to the galaxies themselves, such as
galaxy K or the two interacting galaxies (257 and 255) which are
also a radio source.

As already noticed by \pislar, the wavelet analysis of the HRI image
confirms that the main structure of \a85 has two superimposed X-ray
emitting sources to the west. By combining radio and optical information
we may draw some conclusions on the physical mechanism accounting for
these X-ray sources.  One of them is related to the foreground
structure around the radio Seyfert galaxy A85-F (also noted D in
\pislar). The X-ray excess at this position is therefore probably
produced as a result of nuclear activity in A85-F.  The other X-ray
source is spatially related to radio emission but there is no galaxy
in that location (although there is a foreground group of galaxies).
Thus, if the radio emission is
indeed due to relativistic electrons from a now extinguished active
galaxy, the excess of X-ray emission here is non thermal, probably due
to inverse Compton scattering.

\begin{acknowledgements}

We thank J. Bagchi for enlightening discussions on the radio
emission in clusters.  GBLN acknowledges financial support from the
{\it R\'egion Rh\^one-Alpes, France}. We have also received financial
support from the GDR-Cosmologie, CNRS.

\end{acknowledgements}

\appendix
\section{Appendix: the wavelet image reconstruction}

The noise in the HRI image has been removed according to a multiscale
strategy where only significant structures of different spatial scales
are taken into account to compute a clean image. The key points of
this wavelet-based technique are already given in PDGLS, so that only
the modification included in the second version of the package 
is now discussed. In fact, this minor change from the technical point
of view notably enhances the overall efficiency of the method.

As sketched out in PDGLS and fully described in Ru\'e \& Bijaoui (1997),
objects with a typical size $a$ are identified in the wavelet space from
the corresponding local maxima in the wavelet coefficients for this
spatial scale. Extensive simulations showed indeed that this scheme
yields accurate results for structure detection. However, as usual, the
final test is always provided by the complexity of some images of the
real world. Actually, faint but conspicuous diffuse components appeared
to be missed by the algorithm when much brighter small-scale features
were superimposed onto them, and a second iteration of the program was
therefore necessary to detect such large-scale structures. An example of
such a case occurred in the analysis of the ROSAT/PSPC image of ABCG~85:
the intense peak of X-ray emission found at the location of the central
cD galaxy forced us to run the code twice in order to detect the overall
X-ray emission of the whole cluster of galaxies. So, despite the {\it
a priori} good performances of this two-step method, one might conjecture
that the final restored image could perhaps exhibit some unreal features
caused by reconstruction errors due to this failure. A better solution was 
therefore necessary.

Bijaoui (private communication) solved this lack of sensitivity by
changing the normalisation of the wavelet coefficients when the search
for the local maxima is performed.  The set of functions leading to
the discrete wavelet transform can indeed be normalised so that the
$L^1$-norm or the energy is preserved. Although the latter is surely
more widely used since the translations and dilatations of such an
analysing wavelet define an orthonormal basis of $L^2(R)$, we decided
to choose the former because the wavelet coefficients can thus be
viewed as the difference between true smoothed versions of the
data. But, as explained above, our wavelet-based search for structures
relies on the detection of local maxima along the scale axis, and,
from the image processing point of view, any efficient detection must
involve the energy content of the wavelet coefficient rather than its
amplitude. In fact, once this normalisation with respect to energy has
been introduced for this crucial part of the algorithm, tests show
that most structures are now detected in one step whatever their
relative peak intensities and scales are. So, we made use of this
improved program to remove the noise from the ROSAT images, applying
it now twice only in order to bring out the faintest components.

\end{document}